\documentclass[aps,prd,twocolumn,superscriptaddress,showpacs,preprintnumbers,amsmath,amssymb]{revtex4-2}
\usepackage[utf8]{inputenc}

\usepackage[dvipsnames]{xcolor}
\usepackage[normalem]{ulem}
\usepackage{comment}
\usepackage{overpic}
\usepackage{multirow}
\usepackage{booktabs} 
\usepackage{lineno}

\usepackage[colorlinks,linkcolor=red,anchorcolor=blue,citecolor=blue,urlcolor=blue,breaklinks=true]{hyperref}
\newsavebox{\tablebox}
\usepackage{orcidlink}

\usepackage{graphicx}% Include figure files
\usepackage{dcolumn}% Align table columns on decimal point
\usepackage{bm}% bold math
\usepackage{gensymb}
\usepackage{tabularx}
\usepackage{booktabs}
\usepackage{multirow}
\usepackage{lipsum}
\usepackage{subfigure}

\newcommand{\BESIIIorcid}[1]{\href{https://orcid.org/#1}{\hspace*{0.1em}\raisebox{-0.45ex}{\includegraphics[width=1em]{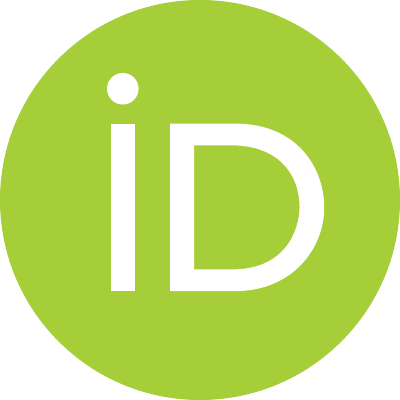}}}}

\begin{document}
\title{Search for the rare decay $\eta^{\prime}\to 3\gamma$} 
\author{
%% Saved at => 2026-04-28
M.~Ablikim$^{1}$\BESIIIorcid{0000-0002-3935-619X},
M.~N.~Achasov$^{4,c}$\BESIIIorcid{0000-0002-9400-8622},
P.~Adlarson$^{84}$\BESIIIorcid{0000-0001-6280-3851},
X.~C.~Ai$^{90}$\BESIIIorcid{0000-0003-3856-2415},
C.~S.~Akondi$^{32A,32B}$\BESIIIorcid{0000-0001-6303-5217},
R.~Aliberti$^{40}$\BESIIIorcid{0000-0003-3500-4012},
A.~Amoroso$^{83A,83C}$\BESIIIorcid{0000-0002-3095-8610},
Q.~An$^{79,66,\dagger}$,
Y.~H.~An$^{90}$\BESIIIorcid{0009-0008-3419-0849},
M.~S.~Anderson$^{40}$\BESIIIorcid{0009-0008-1550-2632}
% ...
\\
\vspace{0.2cm}
(BESIII Collaboration)\\
\vspace{0.2cm} {\it
$^{1}$ Institute of High Energy Physics, Beijing 100049, People's Republic of China\\
% ...
$^{4}$ Budker Institute of Nuclear Physics SB RAS (BINP), Novosibirsk 630090, Russia\\
% ...
$^{32}$ INFN Sezione di Ferrara, (A)INFN Sezione di Ferrara, I-44122, Ferrara, Italy; (B)University of Ferrara, I-44122, Ferrara, Italy\\
% ...
$^{40}$ Johannes Gutenberg University of Mainz, Johann-Joachim-Becher-Weg 45, D-55099 Mainz, Germany\\
% ...
$^{66}$ State Key Laboratory of Particle Detection and Electronics, Beijing 100049, Hefei 230026, People's Republic of China\\
% ...
$^{79}$ University of Science and Technology of China, Hefei 230026, People's Republic of China\\
% ...
$^{83}$ University of Turin and INFN, (A)University of Turin, I-10125, Turin, Italy; (B)University of Eastern Piedmont, I-15121, Alessandria, Italy; (C)INFN, I-10125, Turin, Italy\\
$^{84}$ Uppsala University, Box 516, SE-75120 Uppsala, Sweden\\
% ...
$^{90}$ Zhengzhou University, Zhengzhou 450001, People's Republic of China\\

\vspace{0.2cm}
$^{\dagger}$ Deceased\\
% ...
$^{c}$ Also at the Novosibirsk State University, Novosibirsk, 630090, Russia\\
% ...

}
%% ends here %%

}
\date{\today}

\begin{abstract}
Using a data sample of $(10087 \pm 44) \times 10^{6} J/\psi $ events collected with the BESIII detector, we search for the rare decay $\eta^{\prime}\to 3\gamma$ via the process $J/\psi\to\gamma\eta^{\prime}$. No significant signal is observed and an upper limit on the branching fraction of $\eta^{\prime}\to 3\gamma$ is set at $\mathcal{B}(\eta^{\prime}\to 3\gamma) < 2.25\times 10^{-5}$ at the 90\% confidence level. This result improves the previous upper limit by more than a factor of four.  
%This result improves upon the precision of previous measurements by an order of magnitude.

\end{abstract}

\maketitle
\oddsidemargin -0.2cm
\evensidemargin -0.2cm

%\tableofcontents
%\linenumbers  

\section{\label{sec:introduction}INTRODUCTION}

The neutral pseudoscalar $\eta$ and $\eta^{\prime}$ meson states, 
%formed by light-strange quark mixing, 
arising from the mixing of light-quark SU(3) flavor states,
provide a sensitive probe to test fundamental symmetries and to search for new physics beyond the Standard Model (BSM). Both have relatively small decay widths because many strong and electromagnetic decays are forbidden at leading order due to $P$, $C$, $CP$, $G$-parity, or angular momentum conservation, which enhances the relative importance of new physics signals. Symmetry-violating $\eta, \eta^{\prime}$ decays remain of interest for BSM searches since their decay rates 
are expected to be extremely small within the Standard Model.
%within the Standard Model, induced by the weak interaction, are negligible. 

Searches for $CP$ violation are often conducted through studies of $P$-, $CP$-violating processes. However, limits on the neutron electric dipole moment imply stringent constraints on many other potential signals of such violations. 
On the other hand, the decay $\eta^{\prime}\to 3\gamma$ is mainly studied as a test of $C$ violation, with potential sensitivity to $CP$-violating effects. 
%Consequently, the implications of experiments probing such decays for potential BSM physics are currently unknown, providing substantial research opportunities
Therefore, experimental studies of such decays may provide complementary information on possible BSM scenarios~\cite{GAN20221}.

The decays $\eta, \eta^{\prime}\to 3\gamma$ provide direct tests of $C$ violation. The related channel $\pi^{0} \to 3\gamma$ was initially proposed~\cite{PRENTKI196588} and studied~\cite{BERENDS1965178} for this reason. Since each power of momentum in the amplitude 
introduces a power-law dependence on the pion mass
%leads to a corresponding power of the invariant mass spectrum of the $\pi^{0}$ meson
($m_{\pi^{0}}$) in the decay rate (after phase space integration), the width $\Gamma(\pi^{0}\to3\gamma)$ is suppressed by a very high power of $m_{\pi^{0}}$, namely: 
\begin{equation}
    \frac{\Gamma(\pi^{0}\to 3\gamma)}{\Gamma(\pi^{0}\to 2\gamma)}\approx 1.2\times 10^{-5} \left( \frac{\alpha_{em}G^{2}_{F}m^{4}_{\pi^{0}}}{(2\pi)^{5}} \right) \left(\frac{m_{\pi^{0}}}{\Bar{m}}\right)^{8}\label{eg1},
\end{equation}
where $\Bar{m}$ represents a hadronic scale of $\mathcal{O}$(0.1~-~1 GeV) appearing in the quark loop~\cite{PhysRevD.12.2133}. Here, $\alpha_{em}$ represents the electromagnetic fine-structure constant, and $G_{F}$ represents the Fermi coupling constant. 
%Based on studies of the $\pi^0\to 3\gamma$ decay channel and considering its scaling with the pseudoscalar mass, 
Based on the scaling behavior inferred from studies of $\pi^0\to3\gamma$,
we obtain the Standard Model predictions of branching fractions (BFs)~\cite{GAN20221}.
\begin{gather*}
    \mathcal{B}(\eta\to 3\gamma) \sim 10^{-25}-10^{-17}, \\ 
    \mathcal{B}(\eta^{\prime}\to 3\gamma) \sim 10^{-23}-10^{-15}.
\end{gather*}
Due to low $\eta^{\prime}$ statistics, research on this process remains limited. The current experimental upper limit for the $\eta^{\prime}\to 3\gamma$ BF, $\mathcal{B}(\eta^{\prime}\to 3\gamma) < 1.0\times10^{-4}$ at the 90\% confidence level (C.L.), is based on a result of the GAMS-2000 experiment at Serpukhov~\cite{Serpukhov-Brussels-LosAlamos-AnnecyLAPP:1987kiw}. 
For the $\eta\to 3\gamma$ decay, the most stringent upper limit is $\mathcal{B}(\eta\to 3\gamma) < 1.6\times10^{-5}$ at 90\% C.L., obtained by the KLOE Collaboration in 2004~\cite{KLOE:2004ukf}.

In this paper, we search for the rare decay $\eta^{\prime} \to 3\gamma$, based on an analysis of $10087 \times 10^{6}$ $J/\psi$ events collected with the BESIII detector~\cite{CPC}. The number of produced $\eta^{\prime}$ mesons is estimated to be $(5.33\pm 0.06) \times 10^{7}$, using $\mathcal{B}(J/\psi \to \gamma \eta^{\prime}) = (5.28\pm 0.06) \times 10^{-3}$ from the Particle Data Group (PDG)~\cite{PDG}.  

\section{\label{sec:BESIII DETECTOR AND MONTE CARLO SIMULATION}BESIII DETECTOR AND MONTE CARLO SIMULATION}

The BESIII detector~\cite{Ablikim:2009aa} records symmetric $e^+e^-$ collisions provided by the BEPCII storage ring~\cite{Yu:IPAC2016-TUYA01}
in the center-of-mass energy range from 1.84 to 4.95~GeV, with a peak luminosity of $1.1 \times 10^{33}\;\text{cm}^{-2}\text{s}^{-1}$ 
achieved at $\sqrt{s} = 3.773\;\text{GeV}$. 
Large data samples have been collected in this energy region~\cite{Ablikim:2019hff}. The cylindrical core of the BESIII detector covers 93\% of the full solid angle and consists of a helium-based
 multilayer drift chamber~(MDC), a time-of-flight
system~(TOF), and a CsI(Tl) electromagnetic calorimeter~(EMC),
which are all enclosed in a superconducting solenoidal magnet
providing a 1.0~T magnetic field. The magnetic field was 0.9~T in 2012, which affects 10.7\% of the total $J/\psi$ data.
The solenoid is surrounded by an octagonal flux-return yoke made of steel, interleaved with resistive-plate-counter muon-identification modules.
The charged-particle momentum resolution at $1~{\rm GeV}/c$ is
$0.5\%$, and the 
${\rm d}E/{\rm d}x$
resolution is $6\%$ for electrons
from Bhabha scattering. The EMC measures photon energies with a
resolution of $2.5\%$ ($5\%$) at $1$~GeV in the barrel (end cap)
region. The time resolution in the plastic scintillator TOF barrel region is 68~ps, while that in the end cap region was 110~ps. The end cap TOF system was upgraded in 2015 using multigap resistive plate chamber technology, providing a time resolution of
60~ps, which benefits 87\% of the data used in this analysis~\cite{li2017radiat, guo2017radiat, cao2020nucl}.

Simulated data samples produced with a \textsc{geant4}-based Monte Carlo (MC) package~\cite{geant4}, which includes the geometric description of the BESIII detector and the detector response, are used to determine detection efficiencies and to estimate backgrounds. 
The simulation models the beam-energy spread and initial-state radiation in the $e^{+}e^{-}$ annihilation with the generator {\sc kkmc}~\cite{kkmc2}. 
All particle decays are modeled with {\sc evtgen}~\cite{evtgen_1,evtgen_2} using BF taken from the PDG~\cite{PDG}. The remaining unknown decay modes are modeled with {\sc lundcharm}~\cite{lundcharm}, while other hadronic events are generated with {\sc pythia}~\cite{pythia}.

\section{\label{EVENT SELECTION}EVENT SELECTION AND BACKGROUND ANALYSIS}
\begin{figure}[htbp] 
  \centering
  \includegraphics[width=0.47\textwidth]{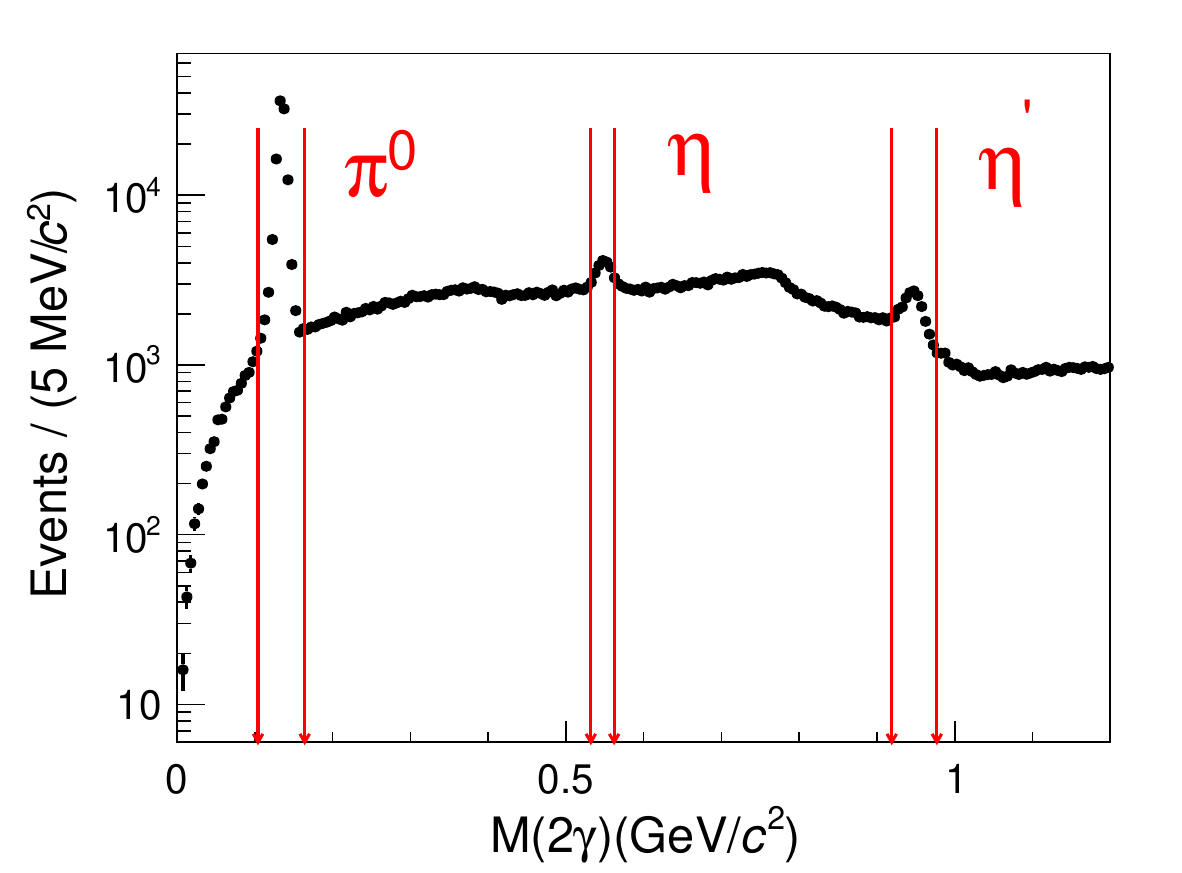}
  
  \caption{The invariant mass distribution of all $2\gamma$ combinations.
  The red arrows, from left to right, indicate the selected mass windows for $\pi^{0}$, $\eta$, and ~$\eta^{\prime}$ decays.}\label{fig:veto}
  \end{figure}

%A search for the decay $\eta^{\prime}\to3\gamma$ has been performed with BESIII among four-photon events, corresponding to the two-step process $J/\psi\to\gamma\eta^{\prime}$, $\eta^{\prime}\to3\gamma$ and the absence of any charged tracks in the final state.
A search for the decay $\eta^{\prime}\to3\gamma$ is performed using events with at least four photons and no charged tracks in the final state, corresponding to the process $J/\psi\to\gamma\eta^{\prime}$ with $\eta^{\prime}\to3\gamma$.
Photon candidates are identified from energy deposit clusters in the EMC. To enhance reconstruction efficiency and energy resolution, energy deposits registered in the adjacent TOF detector are incorporated. To suppress backgrounds arising from low-energy deposits, photon candidates must satisfy minimum energy thresholds: $E > 25\ \mathrm{MeV}$ in the barrel region ($|\cos\theta|<0.80$) and $E > 50\ \mathrm{MeV}$ in the end-cap regions ($0.86<|\cos\theta|<0.92$). Additionally, electronic noise and extraneous energy deposits unrelated to the event are suppressed by applying a timing constraint. The timing difference $\Delta T$, defined as the EMC time of each photon candidate relative to the most energetic photon $(\Delta T = t_{\gamma_i} - t_{\gamma_{\text{max}}})$, is restricted to:~$-500~\mathrm{ns} < \Delta T < 500~\mathrm{ns}$.
To further suppress backgrounds and improve the resolution, a four-constraint (4C) kinematic fit imposing energy and momentum conservation is performed under the $J/\psi\to 4\gamma$ hypothesis. If more than one four-photon combination is available, the one with the smallest $\chi^{2}_{\rm{4C}}$ from the kinematic fit is retained. The $\chi^{2}_{\rm{4C}}$ cut is optimized according to a figure of merit (FOM). 
%According to the Punzi significance~\cite{Punzi:2003bu}, we define the FOM as:~$\frac{\epsilon}{\alpha/2+\sqrt{B}}$, where $\epsilon$ denotes the signal efficiency obtained from signal MC samples and $B$ is the number of background events obtained from background MC samples. $\alpha$ is the desired signal significance in units of standard deviations, refer to the previous $\text{BESIII}$ article, $\alpha$ is set to 3 for rare decay measurements, 
Following the Punzi significance criterion~\cite{Punzi:2003bu}, the FOM is defined as:~$\frac{\epsilon}{\alpha/2+\sqrt{B}}$, where $\epsilon$ is the signal efficiency obtained from signal MC samples and $B$ is the expected number of background events estimated from background MC samples. The parameter $\alpha$ corresponds to the target signal significance, in units of standard deviations.
We set $\alpha=3$ which leads to a requirement of $\chi^{2}_{\rm{4C}} < 14$. 
To reject events containing either three or five photons, 
%we perform a 4C kinematic fit under the $3\gamma$ and $5\gamma$ hypotheses and exclude events for which the fit probabilities of the $3\gamma$ and $5\gamma$ hypotheses are larger than that of the $4\gamma$ hypothesis.
we perform 4C kinematic fits under the $3\gamma$ and $5\gamma$ hypotheses. The fit probability is defined as the $p$-value of the $\chi^2$ distribution, $P(\chi^2, n_{\rm dof}) = \int_{\chi^2}^{\infty} f(z; n_{\rm dof})\,dz$, i.e., the probability that a $\chi^2$ random variable with $n_{\rm dof}$ degrees of freedom exceeds the observed value. We exclude events for which the $3\gamma$ or $5\gamma$ fit probability is larger than that of the $4\gamma$ hypothesis.
Figure~\ref{fig:veto} shows the diphoton invariant mass, $M(\gamma\gamma)$, distribution for all the combinations of the four selected photons. There are obvious $\pi^0$, $\eta$ and $\eta^{\prime}$ signals in the $2\gamma$ invariant mass distributions. 
To suppress backgrounds originating from $\pi^0$, $\eta$, and $\eta^{\prime}$ decays into two photons, the mass vetoes were also optimized using the same FOM. Events are rejected if they satisfy any of the following mass window conditions:
$|M(\gamma\gamma) - m_{\pi^0}| < 0.030~\text{GeV}/c^2,~
|M(\gamma\gamma) - m_{\eta}| < 0.015~\text{GeV}/c^2,~\text{or}~
|M(\gamma\gamma) - m_{\eta^{\prime}}| < 0.029~\text{GeV}/c^2$. 
%Here, $m_{\eta}$ denotes the invariant mass spectrum of the $\eta$ meson, and $m_{\eta^\prime}$ denotes the invariant mass spectrum of the $\eta^\prime$ meson.
Here, $m_{\pi^0}$, $m_\eta$, and $m_{\eta'}$ denote the nominal masses of the $\pi^0$, $\eta$, and $\eta'$ mesons, respectively.
After applying the above selection criteria, the most energetic photon is assumed to originate from the radiative photon in $J/\psi$ decay, and events with a $3\gamma$ invariant mass in the range $0.85$ to $1.05$ GeV/$c^2$ are retained.  

About 10 billion $J/\psi$ inclusive MC events are used to study the potential backgrounds. After imposing all the event selection requirements, 
%the main survived channels in the range $0.85-1.05$ GeV/$c^{2}$ of the invariant mass of 3$\gamma$ from the inclusive MC are shown 
the dominant peaking background channels that survive the selection criteria within the cited $3\gamma$ mass window are summarized in Table~\ref{tab:etap_topo}, where $N_{\rm evt}$ represents the number of normalized events used in the fit. 
%Three peaking backgrounds contribute to the $\eta^{\prime}$ mass region: $J/\psi \to \gamma\eta^{\prime}$, $\eta^{\prime} \to \pi^{0}\gamma\gamma$; $J/\psi \to \gamma\eta^{\prime}$, $\eta^{\prime} \to \pi^{0}\pi^{0}\eta$, $\eta\to2\gamma$  and $J/\psi \to \gamma\eta^{\prime}$, $\eta^{\prime} \to 2\gamma$.
These %dominant peaking 
backgrounds are modeled using exclusive MC shapes in the fit to better estimate their contributions. The remaining background components exhibit a smooth distribution, which is parametrized with a second-order Chebyshev polynomial function.

\begin{table}[htbp]
    \centering
    \caption{
%    The main survived background channels after event selection of $J/\psi\to\gamma\eta^{\prime}, \eta^{\prime}\to 3\gamma$.
Dominant background channels remaining after applying all event selection requirements.
}\label{tab:etap_topo}
    \begin{tabular}{ll}
        \toprule
        \midrule
        Decay  $\hspace{3cm}$ & $N_{\rm evt}$ \\
        \midrule
        $J/\psi \to \gamma\eta^{\prime} , \eta^{\prime} \to \pi^{0} \gamma\gamma$ $\hspace{3cm}$ & 4156 \\
        $J/\psi \to \gamma\eta^{\prime} , \eta^{\prime} \to \pi^{0} \pi^{0} \eta, \eta \to 2\gamma$  $\hspace{3cm}$ & 780 \\
        $J/\psi \to \gamma\eta^{\prime} , \eta^{\prime} \to 2\gamma$ $\hspace{3cm}$ & 319 \\
        \midrule
        \bottomrule
    \end{tabular}

\end{table}

\section{\label{FIT RESULTS}FIT RESULTS}
An unbinned maximum likelihood fit is performed on the $3\gamma$ invariant mass distribution, denoted as $M(3\gamma)$.
The signal probability density function (PDF) is modeled using the signal MC shape convolved with a Gaussian function to account for the mass resolution discrepancy between data and MC simulation.
For the backgrounds, the shapes of the three dominant peaking processes are obtained from MC simulations. Among them, the yields of $\eta^{\prime}\to\pi^{0}\pi^{0}\eta$ and $\eta^{\prime}\to2\gamma$ are fixed according to the expected yields calculated using the PDG BFs, while that of $\eta^{\prime}\to \pi^0\gamma\gamma$ is allowed to float in the fit because of the relatively large uncertainty on its PDG BF.
%The three main peaking backgrounds are fitted with MC shapes, and the $\eta^{\prime}\to\pi^{0}\pi^{0}\eta$ and $\eta^{\prime}\to2\gamma$ yields are fixed (PDG values). However the uncertainty of the branching fraction of $\eta^{\prime}\to \pi^0\gamma\gamma$ in the PDG is very large, so its yield is floated in the fit. The fit result is shown in Fig.~\ref{fig:fit}.
The statistical significance is calculated from the change in the negative log-likelihood value, $-\ln\mathcal{L}$, with and without the signal hypothesis;
the result is 0.45 standard deviations. 

% The significance is derived from the variation of the negative log likelihood function ($-\ln\mathcal{L}$) with signal hypothesis and adjusted fitting degrees of freedom. In particle physics, a signal cannot be claimed when the significance is below five sigma ($5\sigma$).

Since no evident $\eta^{\prime}$ signal is seen in $M(3\gamma)$, a Bayesian method~\cite{PDG} is used to obtain the upper limit of the signal yield at the 90\% C.L.. To do so, the distribution of normalized likelihood values for a series of expected signal yields is taken as the PDF. In obtaining the distribution of likelihood values, the signal of $\eta^{\prime}$ is described by the MC shape function convolved with a Gaussian function, and their parameters are free. The upper limit of the signal yield at the 90\% C.L., denoted as $N_{\text{obs}}^{\text{UL}}$, 
%is defined as the number of events where 90\% of the PDF area lies above zero.
is determined by:
\begin{equation}
    0.9 \int_{0}^{\infty} P(N) \,dN = \int_{0}^{N_{\text{obs}}^{\text{UL}}} P(N) \,dN.
\end{equation}
The upper limit of $\mathcal{B}(\eta^{\prime}\to 3\gamma)$ is calculated as:
\begin{equation}
	\mathcal{B}(\eta^{\prime}\to 3\gamma) < \frac{N_{\rm{obs}}^{\rm{UL}}}{N_{J/\psi} \, \mathcal{B}(J/\psi\to \gamma\eta^{\prime}) \, \epsilon},
\end{equation}
where $N_{J/\psi}$ is the number of $J/\psi$ events in data, $\mathcal{B}(J/\psi\to \gamma\eta^{\prime})$ is the PDG BF, and $\epsilon=(21.40 \pm 0.09) \%$ is the detection efficiency determined from signal MC.
The upper limit on $\mathcal{B}(\eta^{\prime}\to 3\gamma) $ is set to be $ 1.74\times 10^{-5}$ at the 90\% C.L. before incorporating systematic uncertainties.

  \begin{figure}[htbp] 
  \centering
  \includegraphics[width=0.47\textwidth]{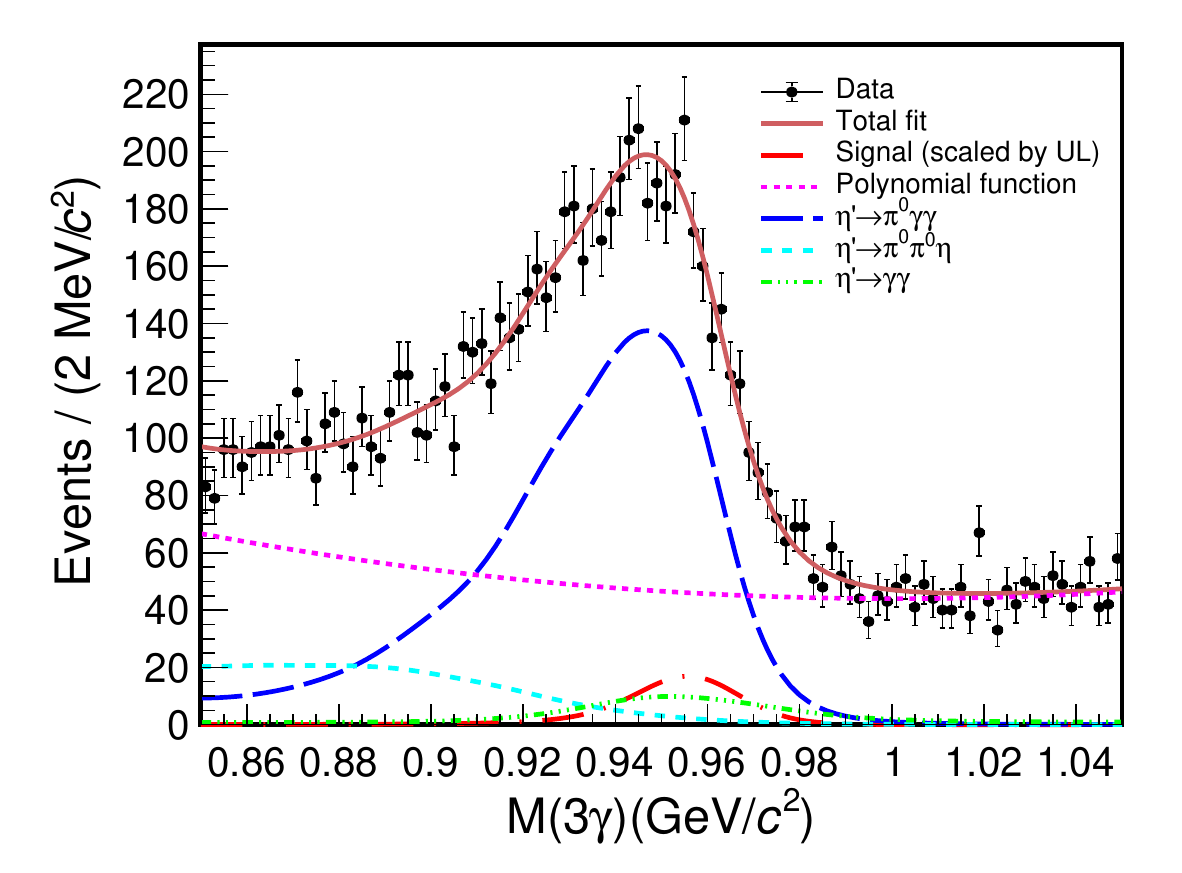}
  \caption{Fit to the $M(3\gamma)$ distribution for $J/\psi\to\gamma\eta^{\prime}, \eta^{\prime}\to3\gamma$ candidates. The black dots with error bars represent data, the brown curve represents the fit result, the red curve represents the signal contribution scaled by upper limit, the dotted pink curve shows the smooth background contribution, the dashed dark blue curve corresponds to the decay  $\eta^{\prime}\to \pi^0\gamma\gamma$, the dashed light blue curve corresponds to the decay  $\eta^{\prime}\to\pi^{0}\pi^{0}\eta$, and the dot-dash green curve corresponds to the decay $\eta^{\prime}\to2\gamma$.}\label{fig:fit}
  \end{figure}
  
\section{\label{sec:SYSTEMATIC UNCERTAINTY}SYSTEMATIC UNCERTAINTY}

The systematic uncertainty considered in the determination of the upper limit is divided into two categories. 
The first category affects the efficiency and is referred to as the multiplicative systematic uncertainty. 
The second category stems from systematic uncertainties in the fitting procedure and is referred to as the additive systematic uncertainty.

The multiplicative systematic uncertainties arise from the number of $J/\psi$ events, photon detection, the cited BFs and the mass window requirements for the $\pi^0, \eta$ and $ \eta^{\prime}$. 
A detailed discussion is provided below:
\begin{itemize}
    \item[(i)]The number of $J/\psi$ events is $(10087\pm 44)\times 10^{6}$, corresponding to an uncertainty of 0.4\%~\cite{CPC}.
    \item[(ii)]The uncertainty of 1.3\% on the BF of $J/\psi\to \gamma\eta^{\prime}$ is taken from the PDG~\cite{PDG}.
    \item[(iii)]The photon efficiency is studied with a clean sample of $J/\psi\to\pi^{+}\pi^{-}\pi^{0}$ decays~\cite{BESIII:2011ysp}. The difference in photon detection efficiency between data and MC simulation, 1\%, is taken as the systematic uncertainty. Since the process involves 4 photons, the total systematic uncertainty from photon detection efficiency is 4.0\%.
    
    \item[(iv)]
    The systematic uncertainties associated with the selection criteria of the $\chi^{2}_{\rm{4C}}$ and the three mass windows ($\pi^0$, $\eta$, and $\eta^{\prime}$) are evaluated using the Barlow test~\cite{Behnke2013}.
    This method quantifies the significance of any deviation through a parameter $\zeta$, defined as:
    \begin{equation}
    \zeta = \frac{V_{\text{nominal}} - V_{\text{test}}}{\sqrt{|\sigma^{2}_{\text{nominal}} - \sigma^{2}_{\text{test}}|}},
    \end{equation}
    to determine whether observed deviations are dominated by statistical fluctuations. Here, ``nominal" denotes the standard analysis configuration: the $\chi^{2}_{\rm{4C}}$ constraint and mass window selections for $\eta$, $\eta^{\prime}$, and $\pi^{0}$. The ``test" configuration corresponds to an alternative configuration with varied $\chi^{2}_{\rm{4C}}$ value or adjusted mass window ranges.
    Multiple Barlow tests are performed, with $V$ representing the signal yield from the fit and $\sigma$ its standard deviation. A smaller $|\zeta|$ value indicates better agreement between nominal and test fits. Systematic uncertainties are considered negligible when $|\zeta| < 2.0$; below this threshold statistical fluctuations dominate differences between configurations. In this analysis, the Barlow tests yield $|\zeta| < 2.0$ for both the $\chi^{2}_{\rm{4C}}$ constraint and the three mass window selections, indicating that the corresponding systematic uncertainties are negligible.

\end{itemize}

Table~\ref{tab:multiplicative} summarizes the multiplicative systematic uncertainties. All sources are treated as uncorrelated, so the total systematic uncertainty is obtained by summing them in quadrature.

\begin{table}[htbp]

    \centering
    \caption{The relative multiplicative systematic uncertainties (\%).}\label{tab:multiplicative}
    \scalebox{0.95}{
    \begin{tabular}{ll}
        \toprule
        \midrule
        Source $\hspace{3cm}$& Uncertainty \\
        \midrule
        $N_{J/\psi}$ $\hspace{3cm}$& 0.4 \\
        $\mathcal{B}(J/\psi \to \gamma\eta^{\prime})$ $\hspace{3cm}$& 1.3 \\ 
        Photon detection $\hspace{3cm}$& 4.0 \\
        $\chi^{2}_{\rm{4C}}$ $\hspace{3cm}$& negligible \\
        $\pi^0,\eta,\eta^{\prime}$ mass windows $\hspace{3cm}$& negligible \\
        \midrule
        Total $\hspace{3cm}$& 4.2\\
        \midrule
        \bottomrule
        
    \end{tabular}
    }
\end{table}

The additive uncertainties arise from the fit range, signal shape, non-peaking background shape and the number of peaking background events. 
We perform alternative fits using the following approaches, adopting the largest result among these fits as a conservative estimation of the systematic uncertainty on the BF upper limit:
\begin{itemize}
    \item[(i)]The systematic uncertainty arising from the fit range is estimated by varying the fit range of $M(3\gamma)$ by $\pm10$ MeV/$c^2$.
    \item[(ii)]The uncertainty due to the signal shape is estimated by fitting the signal PDF without the Gaussian convolution.
    \item[(iii)]The uncertainty due to the non-peaking background shape is estimated by replacing the second-order Chebyshev polynomial with a third-order Chebyshev polynomial.
    \item[(iv)] The uncertainty from peaking backgrounds is estimated by treating the yields of the three background components as free parameters, whereas only the yield of $\eta^{\prime}\to\pi^{0}\gamma\gamma$ is free in the nominal fit.
    
    The resulting variation is ascribed to the corresponding systematic uncertainty. We further note that the results from this unconstrained fit remain consistent with the PDG values.

\end{itemize}

To incorporate the multiplicative systematic uncertainties in the calculation of the upper limit, the likelihood distribution is smeared by a Gaussian function with a mean of zero and a width equal to $\sigma_{\epsilon}$~\cite{2006physics...5236S,Liu_2015}
\begin{equation}
    L'(n')\propto\int_{0}^{\infty} L\left( n\frac{\epsilon}{\epsilon_0} \right) \, {\rm exp} \left[\frac{-{(\epsilon - \epsilon_0)}^{2}}{2\sigma_{\epsilon}^{2}}\right]d\epsilon,
\end{equation} 
where $L(n) \, (L'(n'))$ is the likelihood distribution as a Gaussian function of the yield $n \, (n')$, $\epsilon_0$ is the detection efficiency and $\sigma_{\epsilon}$ is the multiplicative systematic uncertainty.
The upper limit on the 
signal yield
%number of $\eta^{\prime}\to 3\gamma$ events
, $N_{\rm{UL}}'$, is determined to be 255.3, and the corresponding upper limit on $\mathcal{B}(\eta^{\prime}\to 3\gamma) $ is set to be $ 2.25\times 10^{-5}$ at the 90\% C.L..

\section{\label{sec:SUMMARY}SUMMARY}
The rare decay $\eta^{\prime} \to 3\gamma$ is searched for via $J/\psi \to \gamma\eta^{\prime}$ by analyzing a data sample of 10 billion $J/\psi$ events collected with the BESIII detector. No significant $\eta^{\prime}$ signal is observed, and an upper limit on the BF of $\eta^{\prime}\to3\gamma$ is determined to be $\mathcal{B}(\eta^{\prime}\to 3\gamma) < 2.25\times 10^{-5}$ at the 90\% C.L.. This result represents the most stringent upper limit to date,
improving the previous upper limit by more than a factor of four and providing more stringent experimental constraints on $C$-violating $\eta^{\prime}$ decays, with potential implications for searches for physics beyond the Standard Model, including possible new sources of $CP$ violation. 
% with an order of magnitude improvement in precision compared to previous results, which provides a stronger experimental constraint for direct searches of $CP$ violation involving $C$-violating processes.

\section*{\label{sec:ACKNOWLEDGEMENT}ACKNOWLEDGEMENT}

The BESIII Collaboration thanks the staff of BEPCII (https://cstr.cn/31109.02.BEPC) and the IHEP computing center for their strong support. This work is supported in part by National Key R\&D Program of China under Contracts Nos. 2025YFA1613900, 2023YFA1606000, 2023YFA1606704; National Natural Science Foundation of China (NSFC) under Contracts Nos. 11635010, 11935015, 11935016, 11935018, 11905037, 12025502, 12035009, 12035013, 12061131003, 12192260, 12192261, 12192262, 12192263, 12192264, 12192265, 12221005, 12225509, 12235017, 12342502, 12361141819, 12535005; the Chinese Academy of Sciences (CAS) Large-Scale Scientific Facility Program; the Strategic Priority Research Program of Chinese Academy of Sciences under Contract No. XDA0480600; CAS under Contract No. YSBR-101; 100 Talents Program of CAS; The Institute of Nuclear and Particle Physics (INPAC) and Shanghai Key Laboratory for Particle Physics and Cosmology; Agencia Nacional de Investigación y Desarrollo de Chile (ANID), Chile under Contract No. ANID CCTVal CIA250027; ERC under Contract No. 758462; German Research Foundation DFG under Contract No. FOR5327; Istituto Nazionale di Fisica Nucleare, Italy; Knut and Alice Wallenberg Foundation under Contracts Nos. 2021.0174, 2021.0299, 2023.0315; Ministry of Development of Turkey under Contract No. DPT2006K-120470; National Research Foundation of Korea under Contract No. RS-2026-25486791; National Science and Technology fund of Mongolia; Polish National Science Centre under Contract No. 2024/53/B/ST2/00975; STFC (United Kingdom); Swedish Research Council under Contract No. 2019.04595; U. S. Department of Energy under Contract No. DE-FG02-05ER41374

\bibliographystyle{apsrev4-2}
\bibliography{reference}% Produces the bibliography via BibTeX.

\end{document}